\documentclass[12pt,linenumbers]{article}
\usepackage{graphics}
\usepackage{amsmath}
\usepackage{epsfig}
\usepackage{cite}
\usepackage{lineno}
\usepackage[varg]{txfonts}
\newcommand{\be}{\begin{eqnarray}}
\newcommand{\ee}{\end{eqnarray}}
\title{Ratio of kaon-to-pion production cross-sections in
$Be Be$ collisions as a function of $\sqrt{s}$}

\author{G.I.~Lykasov, A.I.~Malakhov, A.A.~Zaitsev} 
\begin{document}
\date{}
\maketitle

\begin{center}

{Joint Institute for Nuclear Research, Dubna 141980, Moscow region, Russia

\vspace{0.5cm}

%artemenkov.denis@gmail.com\\
lykasov@jinr.ru\\
malakhov@lhe.jinr.ru\\
zaicev@jinr.ru\\
}

\end{center}

\vspace{0.5cm}

\begin{center}

{\bf Abstract }

\end{center} 

%\indent
%{\bf
The inclusive spectra of pions and kaons produced in $Be Be$ collisions as functions of their 
transverse momentum $p_t$ at mid-rapidity are calculated within the self-similarity approach.  
A satisfactory description of NA61/SHINE data on these spectra and the
ratio of $K^\pm$ to $\pi^\pm$ meson production cross sections in $Be Be$ collisions as a function of
$\sqrt{s}$ are presented. The similarity of these observables to the ones for $pp$
collisions at mid-rapidity and in the wide range of initial energies is illustrated
%}
     
\vspace{1.0cm}

\noindent
%PACS number(s): 12.15.Ji, 12.38.Bx, 13.85.Qk

%\newpage
\indent
%%%%%%%%%%%%%%%%%%%%%%%%%%%%%%%%%%%%%%%%%%%%%%%%%%%%%%%%%%%%%%%%%%%%%%%%%%%%%%%%%%%%%%%%%%%%%%%%%%%%%%
\section{Introduction}
\label{intro}
%{\bf
The investigation of strange hadron and pion production in heavy-ion collisions
is a promising tool to search for new physical properties of such processes.
The observation of a sharp peak in the production ratio of $K^+$ 
mesons  to $\pi^+$ mesons in central $Pb+Pb$ and $Au+Au$ collisions at mid-rapidity 
\cite{NA49:2002,NA49:2008} has attracted the attention of both theoreticians and 
experimentalists see \cite{Marek:1999}-\cite{NA61/SHINE:2020} and references therein. 
When the initial energy $\sqrt{s_{NN}}$ per nucleon becomes larger than 30 GeV this 
ratio falls down.
%%%%%%%%%%%%%%%%%%%%%%%%%%%%%%%%%%%%%%%%%%%%%%%%%%%%%%%%%%%%%%%%%%%%%%%
However, the ratio of $K^+$ to $\pi^+$ mesons produced in collisions 
of nuclei lighter than $Pb$ and $Au$, as a function of the initial energy
$\sqrt{s_{NN}}$ per nucleon, in particular, $Be+Be$ \cite{NA61_BeBe:2021} 
and $Ar+Sc$ \cite{NA61_ArSc_ratio:2021} has no peak. The fast increase of this 
ratio, when $\sqrt{s_{NN}}$ grows from the kaon threshold up to 20-30 GeV and the 
slow increase at larger energies has been observed 
\cite{NA61_BeBe:2021,NA61_ArSc:2021}. A similar energy dependence 
is observed by the NA61/SHINE Collaboration in $Be Be$ and 
$Ar Sc$ collisions. Moreover, the energy dependence of $K/\pi$ ratios observed 
in $Be Be$ collisions \cite{NA61_BeBe:2021} is similar to the one 
in $p p$ collisions.

%%%%%%%%%%%%%%%%%%%%%%%%%%%%%%%%%%%%%%%%%%%%%%%%%%%%%%%%%%%%%%%%%%%%%%%%%%%%%%%%%%%%%

In this paper we analyze the production of kaons and pions in $Be Be$ collisions 
at mid-rapidity and focus on ratios between their cross-sections as functions of
the initial energy within the same theoretical approach, which was presented 
in \cite{LMZ:2021} for $pp$ collisions. This approach is based on the similarity of 
spectra of hadrons produced in $AA$ collisions at zero rapidity $y=0$
and on the conservation laws of four-momenta and quantum numbers
suggested in \cite{c3,c4,4,5}. It was a continue of the approach suggested earlier  
in pioneering papers \cite{Fermi:1950,Pomeran:1951,Landau:1953,Hagedorn:1965}.
Further development of this approach was presented in 
\cite{Baldin_AA:1996,ALM:2015,LM:2018,ML:2020,LMZ:2021}.  

%}

\section{Main properties of the self-similarity approach for $A A$ collisions.}
\label{sec:1}
The inclusive production of hadron $1$ in the interaction of nucleus $A$ with nucleus $B$     
\begin{equation}
A  +  B  \rightarrow  1  + \ldots ,                                           
\label{eq:n1}                                                                
\end{equation}                                              
is satisfied by the conservation law of four-momenta in the following form \cite{4,5}:
\begin{equation}
{(N_AP_A + N_BP_{B} - p_1)}^2 = 
{(N_Am_0 + N_B m_0 + M)}^2 ,
\label{eq:n2}
\end{equation}
where $N_A$ and $N_B$ are the fractions of the four-momentum transmitted by
nucleus $A$ and nucleus $B$, the forms of $N_A, N_B$  are presented in \cite{5,LM:2018} ; 
$P_A$ , $P_B$ , $p_1$ are the four-momenta of nuclei  
$A$ and $B$ and particle $1$,   respectively; $m_0$ is the mass of the free nucleon; $M$ is
the mass of the 
particle providing for conservation of the baryon 
number, strangeness, and other quantum numbers.
Eq.~\ref{eq:n2} was introduced in \cite{4,5} for the production of hadrons in
$AB$ collisions in the kinematics 
forbidden 
for free nucleon-nucleon collisions. In fact, it is valid for initial energies
of colliding nuclei close to the threshold of hadron production.  
It allows us to find the minimal value of $M$,
which provides for the conservation of quantum numbers.   
For $\pi$-mesons $m_1 = m_\pi$  and $M = $0.
For anti nuclei $M=m_1$ and for $K^-$-mesons  $M = m_1 = m_K$,
$m_K$ is the mass of the $K$-meson.
For nuclear fragments $M = - m_1$.
For $K^+$-mesons $m_1 = m_K$ and $M = m_\Lambda  - m_0$,
$m_\Lambda$ is the mass of the $\Lambda$-baryon.
Let us note that the isospin effects of the produced hadrons and other nuclear effects
are out of this approach. Therefore, it is assumed that within the self-similarity
approach there is no big
difference between the inclusive spectra of $\pi^+$ and $\pi^-$ mesons produced in $pp$ and $AA$
collisions. However, there is a difference between similar spectra of $K^+$ and $K^-$ mesons, because 
the values of $M$ are different. This is due to the conservation law of strangeness.

 In \cite{4,5} the parameter of self-similarity is introduced in the following 
form:
\begin{equation} 
\Pi=\min \frac{1}{2} \left[ (u_A N_A + u_B N_B)^2\right]^{1/2}  ,
\label{eq:n3} 
\end{equation}                                            
where $u_A$ and $u_B$ are the four-velocities of nuclei $A$ and $B$. 
The minimization over 
$N$ presented in Eq.~(\ref{eq:n3}) allows us to find the parameter $\Pi$. This
parameter introduced in \cite{4} was obtained in \cite{5} for nucleus-nucleus
collisions in the mid-rapidity region, however, it can
also be applied successfully for the analysis of 
pion production in $pp$ collisions, as it was shown in
\cite{ALM:2015,LM:2018,ML:2020}. 

The inclusive spectrum of particle 1 produced in the $AB$ collision
can be parameterized as a general universal function dependent on the 
self-similarity parameter $\Pi$, as it was shown in \cite{Baldin_AA:1996}.
\begin{equation}
E d^3 \sigma_{AB}/d^3p~=~A_A^{\alpha(N_A)}\cdot A_B^{\alpha(N_{B})}\cdot F(\Pi)
%\exp(-\Pi/C_2),
\label{eq:n4} 
\end{equation}
where $\alpha(N_A)=1/3 + N_A/3$, $\alpha(N_B)=1/3 + N_B/3$ 
and function $F(\Pi)$ is the inclusive spectrum of hadron production in the $NN$ collision 
\cite{LM:2018,ML:2020}, $A_A$ and $A_B$ is the number of nucleons in nuclei $A$ and $B$,
respectively.

The form of Eq.~\ref{eq:n4} allowed us to describe satisfactorily 
inclusive spectra of hadrons produced in $pA$ and $AB$ collisions in 
kinematics forbidden at hadron production in $pp$ interactions and of 
particles produced close to the threshold of nucleon-nucleon collisions.  
This form of the inclusive hadron spectrum also results in the satisfactory 
description of $p_t$ spectra of pions produced
in $AB$ collisions at the mid-rapidity region and not large
transverse momenta $p_t$ of produced pions in the wide range of initial
energies \cite{ALM:2015,LM:2018,AJLLM:2018,ML:2020}.
Therefore, we apply it to describe $p_t$ spectra of kaons produced in
$AB$ collisions at mid-rapidity and not large transverse momenta.  
     
For symmetric colliding nuclei $N_A=N_B=N$ the function $\Pi$ is found from 
the minimization of Eq.~\ref{eq:n3} by solving the equation. This assumption 
has been suggested in \cite{4,5} 
\begin{eqnarray}
\frac{d\Pi}{d N}=0
\label{def:minimPi} 
\end{eqnarray}  
The exact solution of Eq.~\ref{def:minimPi} at zero rapidity $y=0$,
as $N=\Pi/cosh(Y)\equiv 2m_0\Pi/\sqrt{s}$, 
was obtained in \cite{5}, for details see, also \cite{LM:2018}.  
Therefore, $\alpha(N)=1/3 + 2m_0\Pi/(3\sqrt{s})$.
For symmetric nuclei Eq.~\ref{eq:n4} is presented in the following form
\begin{equation}
E d^3 \sigma_{AA}/d^3p~=~A^{2\alpha(N)}\cdot F(\Pi)
\label{eq:AA_sp} 
\end{equation} 
\begin{eqnarray}
F(\Pi)=\bigg[ A_q \mbox{exp}\Big(-\frac{\Pi}{C_q}\Big) + \\
\nonumber
A_g\sqrt{p_T}\phi_1(s) \mbox{exp}\Big(-\frac{\Pi}{C_g}\Big)\bigg] \sigma_{tot}
\label{def:F} 
\end{eqnarray}
where 
\begin{eqnarray}
\Pi(s,m_{1T},y)~=~\left\{\frac{m_{1T}}{2m_0\delta_h}+
\frac{M}{\sqrt{s}\delta_h}\right\}\mbox{cosh}(y)G ,
\label{eq:n10} \\
\nonumber
G = \left\{1+\sqrt{1+\frac{M^2-m_1^2}
{(m_{1T}+2Mm_0/\sqrt{s})^2\mbox{cosh}^2(y)}\delta_h}\right\}~.
\end{eqnarray}
Here 
$\phi_1(s)~=~1-\sigma_{nd}(s)/\sigma_{tot}(s)$, see\cite{LM:2018,ML:2020},\\
$\delta_h=\left(1 - \frac{s_{th}^h}{s} \right)$;
$s_{th}^{\pi}\simeq 4m_0^2$; 
$s_{th}^{K^+}=\left(m_0 + m_K + m_\Lambda \right)^2$;
$s_{th}^{K^-}=(2m_0+2m_K)^2$;
$M = m_\Lambda - m_0; m_\Lambda = $ 1.115 GeV; 
$m_k = $ 0.494 GeV; $s_0 = $ 1 GeV; $m_0 = 0.938$ GeV;
$p_{1T}$ and $m_{1T}$ are the transverse momentum and transverse mass
of the produced hadron $1$; 
$\sigma_{nd} = (\sigma_{tot} - \sigma_{el} - \sigma_{SD})$ is the 
non-diffractive cross-section;
$\sigma_{tot},\sigma_{SD}$ and $\sigma_{el}$ are the total
cross-section, the single diffractive cross section and the elastic
cross-section of $pp$ collisions, respectively. They were taken from
\cite{sigma:2013} and \cite{sigm_el:2017} and, together with
 parameters $A_q, C_q$ and  $A_g, C_g$, they are presented in the Appendix.

%{\bf
Note, the self-similarity parameter $\Pi$ depends not only on the transverse mass $m_{1T}$ 
and the rapidity $y$ of the produced hadron but also on the initial energy $\sqrt{s}$.
It leads to the non factorized form of the inclusive spectrum presented as 
Eqs.~(\ref{eq:n4},\ref{eq:AA_sp}), as a function of $s, m_T,y$. At large $\sqrt{s}$ the energy
dependence of $\Pi$ vanishes. This is the main difference of this approach from another models.
%}

In fact, the function $F(\Pi)$ in Eq.~(7) is the inclusive spectrum of hadrons produced in $pp$ 
collisions at the mid-rapidity ($y\simeq$ 0) , which was calculated within the
approach suggested in \cite{BGLP:2012,GLLZ:2013}. The form of  
$F(\Pi)$ is due to the quark contribution (the first term of Eq.~(7)) and the
contribution of nonperturbative gluons (the second term
of Eq.~(7)). The forms of $F(\Pi)$ and $E d^3 \sigma_{AA}/d^3p$  allowed us to
describe satisfactorily the data on inclusive spectra of hadrons 
produced in $pp$ and $AA$ collisions at $y\simeq$ 0 and the wide range of the
initial energies $\sqrt{s}$ \cite{ALM:2015,LM:2018,ML:2020,LMZ:2021}.   
From the best description of the inclusive spectrum of charged 
hadrons produced in $pp$ collisions at LHC energies the transverse momentum 
dependence on the gluon density (TMD) at low square transfers $Q^2$ was constructed in
\cite{GLLZ:2013}. The use of it allowed us to describe rather satisfactorily LHC data
on hard $pp$ processes, HERA data on deep inelastic scattering (DIS) 
and ZEUS data on the charmed and bottom structure functions $F_{2c},F_{2b}$
\cite{LLZ:2014,AJLLM:2018}. Therefore, we can quite reasonably apply $F(\Pi)$ 
in the form of Eq.~(7) entered into Eq.~\ref{eq:AA_sp} to analyze the $p_t$
spectra of hadrons produced in $AA$ collisions at the mid-rapidity range at not large $p_t$.
     
%}

\section{Results and discussion}

In the case of the process $AA\rightarrow h + X$ Eq.~\ref{eq:n4} looks
as the following:
\begin{eqnarray}
\rho_{AA}^h(p_{hT},y)~\equiv E_h\frac{d^3 \sigma^h_{AA}}{d^3p_1}~=
\frac{1}{\pi}\frac{d\sigma^h_{AA}}{dp_{1T}^2dy}~= \\
\nonumber
\frac{1}{\pi}\frac{d\sigma^h_{AA}}{dm_{1T}^2dy}=A^{2\alpha(N)}
F(\Pi(s,m_{1T},y)) ,
\label{eq:AA_yspectrum}
\end{eqnarray}
Then, the production cross-section of  hadron $h$ in $AA$ collisions
integrated over its
transverse momentum $p_{1T}$ or transverse mass $m_{1T}$ at zero rapidity
$y=0$ and $s\geq s_{th}^h$ can be presented in the following form:
\begin{eqnarray}
	\frac{d\sigma^h_{AA}}{dy}(s,y=0)=
	2\pi\int_{p^{\rm{min}}_{1T}}^{p^{\rm{max}}_{1T}}\rho_{h_1}^{AA}(s,p_{1T},y=0)p_{1T}dp_{1T}
\end{eqnarray}
Then, the ratio of production cross-sections for
hadrons $h_1$ and hadrons $h_2$ 
in $AA$ collisions and at $y=0$ as $\sqrt{s}$ can
be presented in the following form:
\begin{eqnarray}
R_{AA}^{h_1h_2}=\frac{d\sigma_{AA}^{h_1}}{dy}(s,y=0)/\frac{d\sigma_{AA}^{h_2}}{dy}(s,y=0)
\label{eq:ratio_AA_crsec}
\end{eqnarray}   
 
\begin{figure}[h] 
\begin{center}
\includegraphics[width=0.9\textwidth]{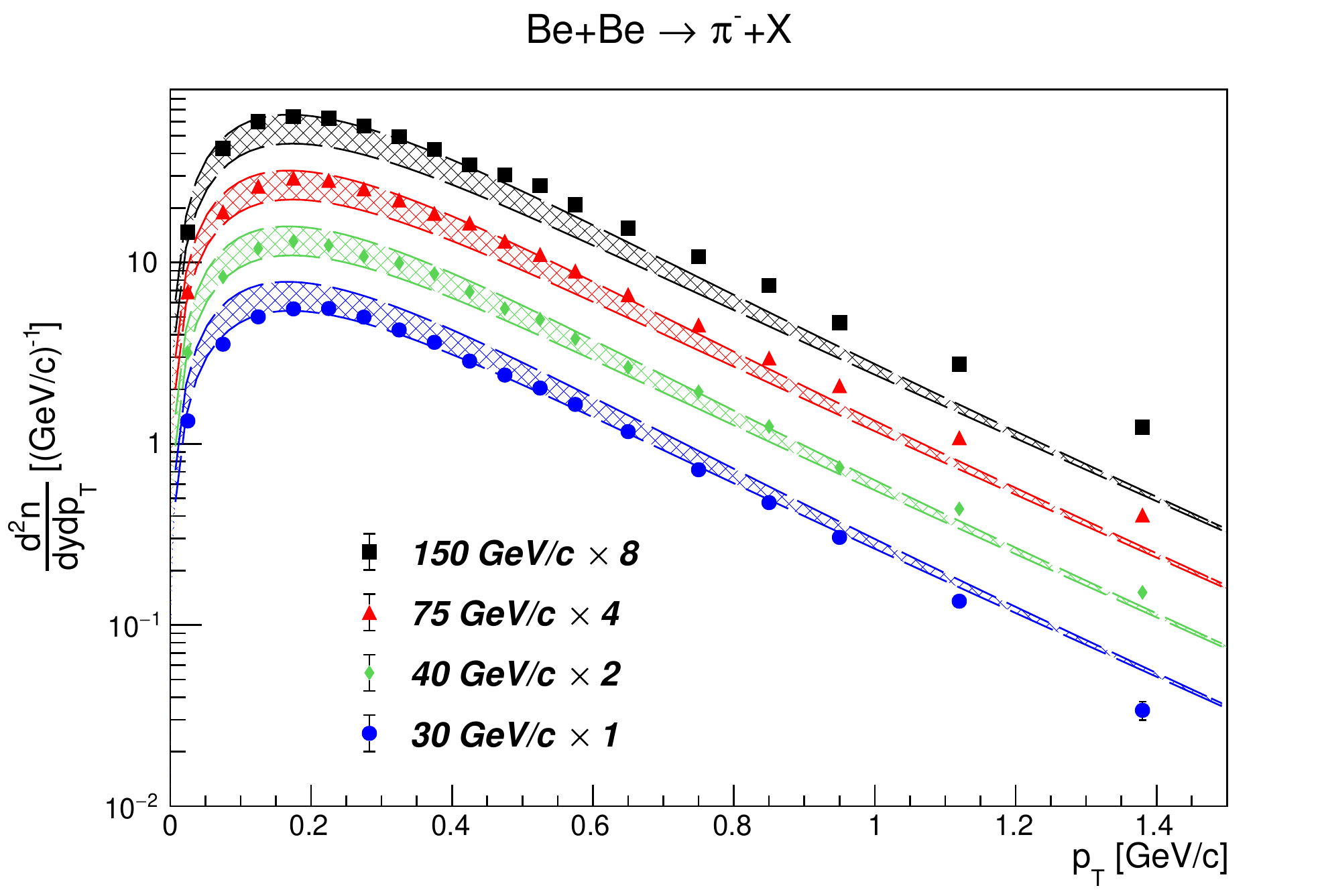}
\end{center}
 \caption{
The $p_T$-spectra of $\pi^-$ mesons produced in the 20\% most central $Be Be$
collisions at momenta of the initial $Be$ nucleus $P_{in}=$
150$A$ GeV$/$c ($\sqrt{s}$=16.84 GeV per nucleon),
75$A$ GeV$/$c ($\sqrt{s}$ = 11.94 GeV per nucleon), 40$A$  GeV$/$c 
($\sqrt{s}$ = 8.77 GeV per nucleon), 30$A$ GeV$/$c ($\sqrt{s}$ = 7.68 GeV per nucleon) 
at mid-rapidity $y<$ 0.2. The lines are
our calculations, the data are taken from 
\cite{NA61_BeBe:2020} and normalized to 20\% of centrality. The bands are due to
uncertainties in parameter $A_q$ presented in the Appendix.
}  
\label{fig_pimin}
\end{figure}
\begin{figure}[h] 
\begin{center}
\includegraphics[width=0.9\textwidth]{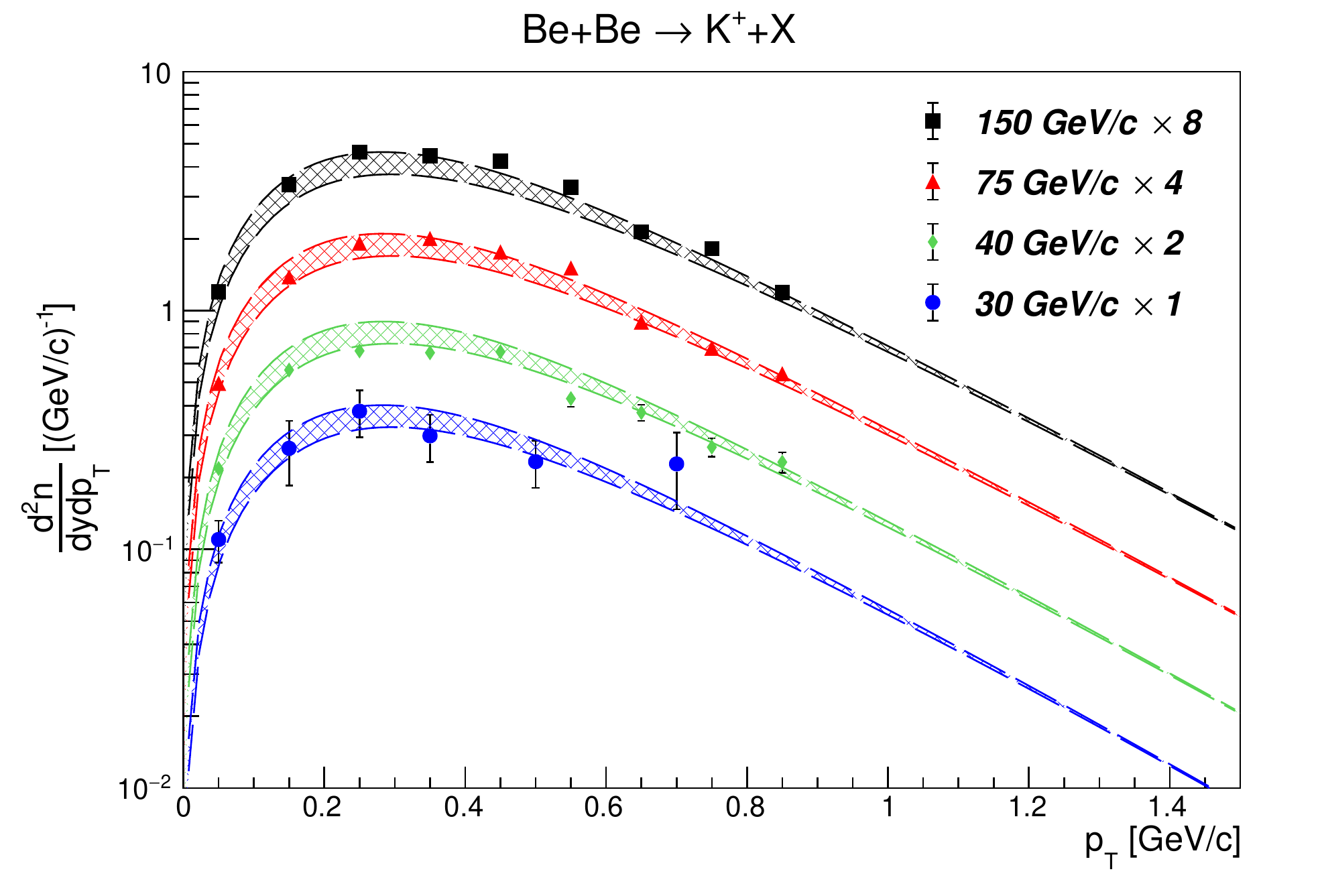}
\end{center}
 \caption{
The $p_T$-spectra of $K^+$ mesons produced in central $BeBe$ collisions.
Notations are the same as in Fig.~\ref{fig_pimin}. The NA61/SHINE data were 
taken from \cite{NA61_BeBe:2021}.
}  
\label{fig_Kplus}
\end{figure}
%%%%%%%%%%%%%%%%%%%%%%%%%%%%%%%%%%%%%%%%%%%%%%%%%%%%%%%%%%%%%%%%%%%%%%%%%%%%%%%%%%%
\begin{figure}[h] 
\begin{center}
\includegraphics[width=0.9\textwidth]{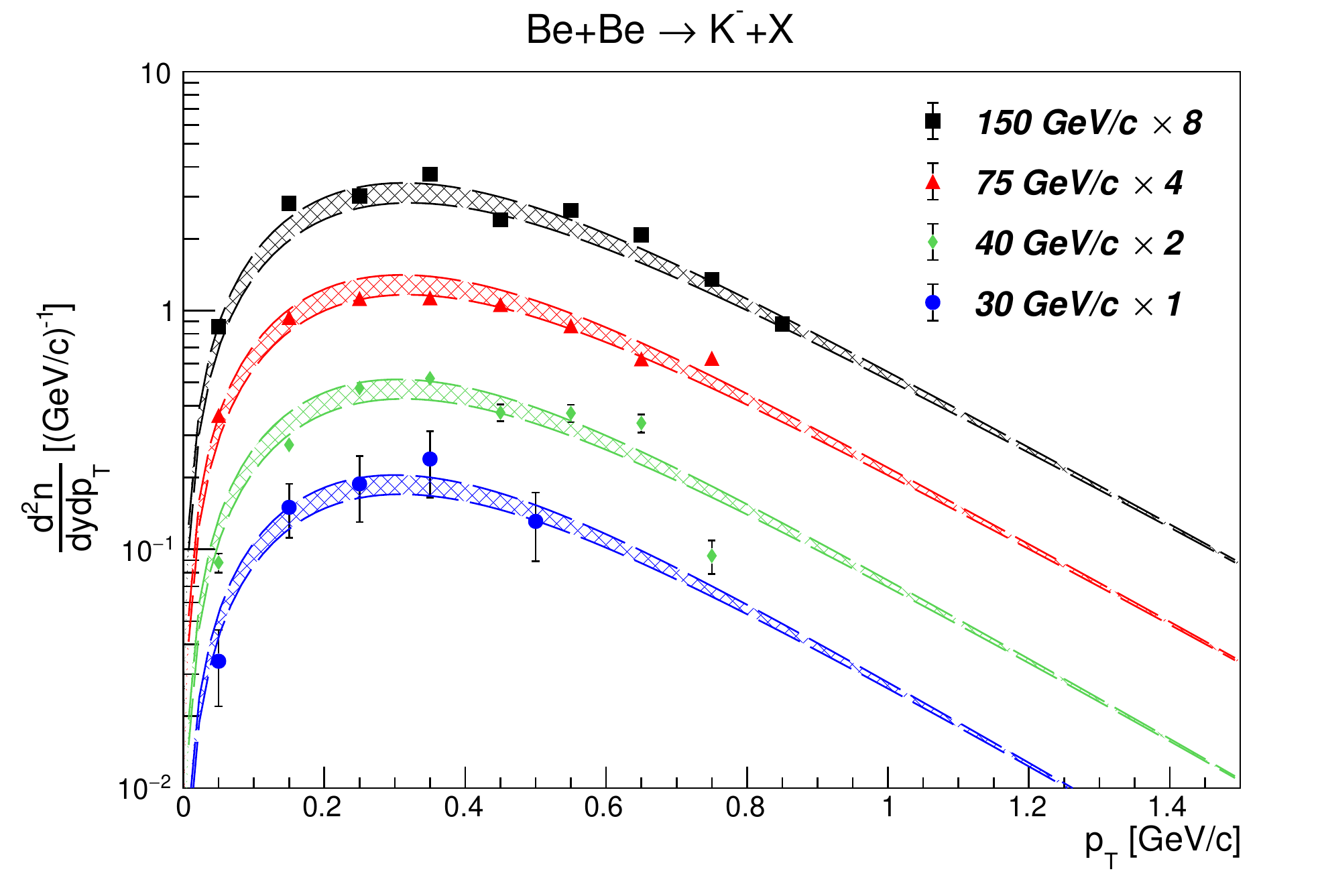}
\end{center}
 \caption{
The $p_T$-spectra of $K^-$ mesons produced in central $BeBe$ collisions.
Notations are the same as in Fig.~\ref{fig_pimin}. The NA61/SHINE data were
taken from \cite{NA61_BeBe:2021}.
}  
\label{fig_Kmin}
\end{figure}
The $p_T$-spectra of $\pi^-$, $K^+$ and $K^-$ mesons are the sums
of quark and gluon contributions including uncertainties due to the fit of data that are
presented in Figs.~(\ref{fig_pimin}-\ref{fig_Kmin}). 
%%%%%%%%%%%%%%%%%%%%%%%%%%%%%%%%%%%%%%%%%%%%%%%%%%%%%%%%%%%%%%%%%%%%%%%%%%%%%%%
By fitting NA61/SHINE data on $p_T$-spectra at mid-rapidity the parameters
$C_q,A_g,C_g$ were found to be independent of the initial energy $\sqrt{s}$,
they depend on the kind of mesons produced, $\pi,K^+,K^-$.
However, the parameter $A_q$ varies a little bit at energies between
40$A$ GeV$/$c and 150$A$ GeV$/$c. The uncertainties in $p_T$-spectra and
ratios of yields, $K^+/\pi^+$ and $K^-/\pi^-$, are due to the uncertainties in
the parameter 
$A_q$. All these parameters are presented in the
Appendix. Similar spectra with quark and gluon contributions are also
presented in the Appendix.

In Figs.~(\ref{fig_ratKplpipl},\ref{fig_ratKmin_20}) the respective yield
ratios, $K^+/\pi^+$ and $K^-/\pi^-$, are presented as functions 
of $\sqrt{s}$. 
From these figures one can see their fast rise from the threshold energy of
$K^+$ or $K^-$ production up to $\sqrt{s}=$ 20-30 GeV and
their further slow increase with energy. 
\begin{figure}[h] 
\begin{center}
\includegraphics[width=0.9\textwidth]{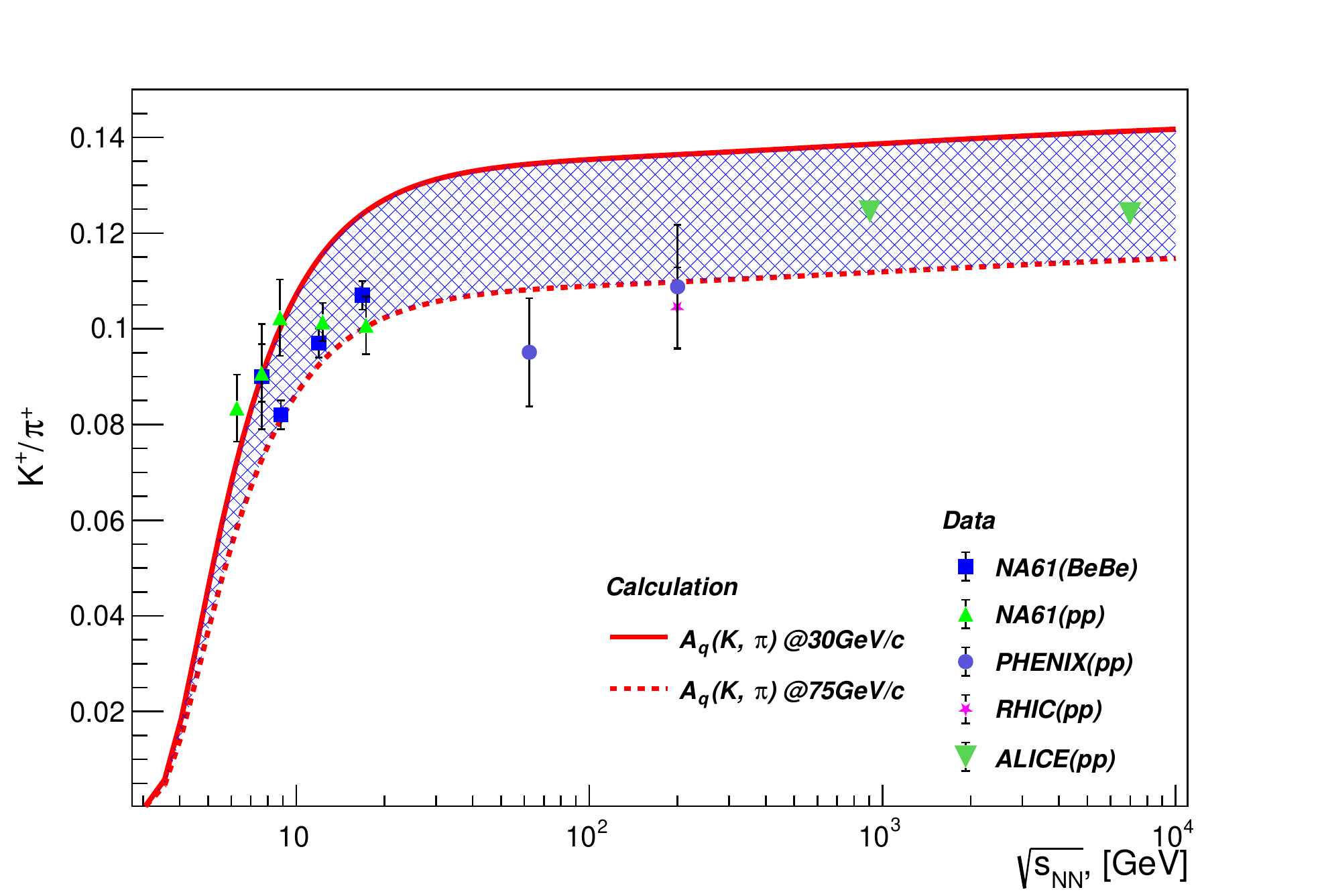}                
\end{center}
 \caption{The ratio of $K^+$ and $\pi^+$ meson yields produced in the
 mean-rapidity of 
$BeBe$ (NA61/SHINE) \cite{NA61_BeBe:2021} and $pp$ (NA61/SHINE,PHENIX,STAR,ALICE) \cite{NA61_pp:2020,STAR_pp:2011,PHENIX_pp:2011,ALICE_pp:2011,ALICE_pp:2015}
collisions as a function of $\sqrt{s}$.
 }  
\label{fig_ratKplpipl}
\end{figure}

\begin{figure}[h] 
\begin{center}
\includegraphics[width=0.9\textwidth]{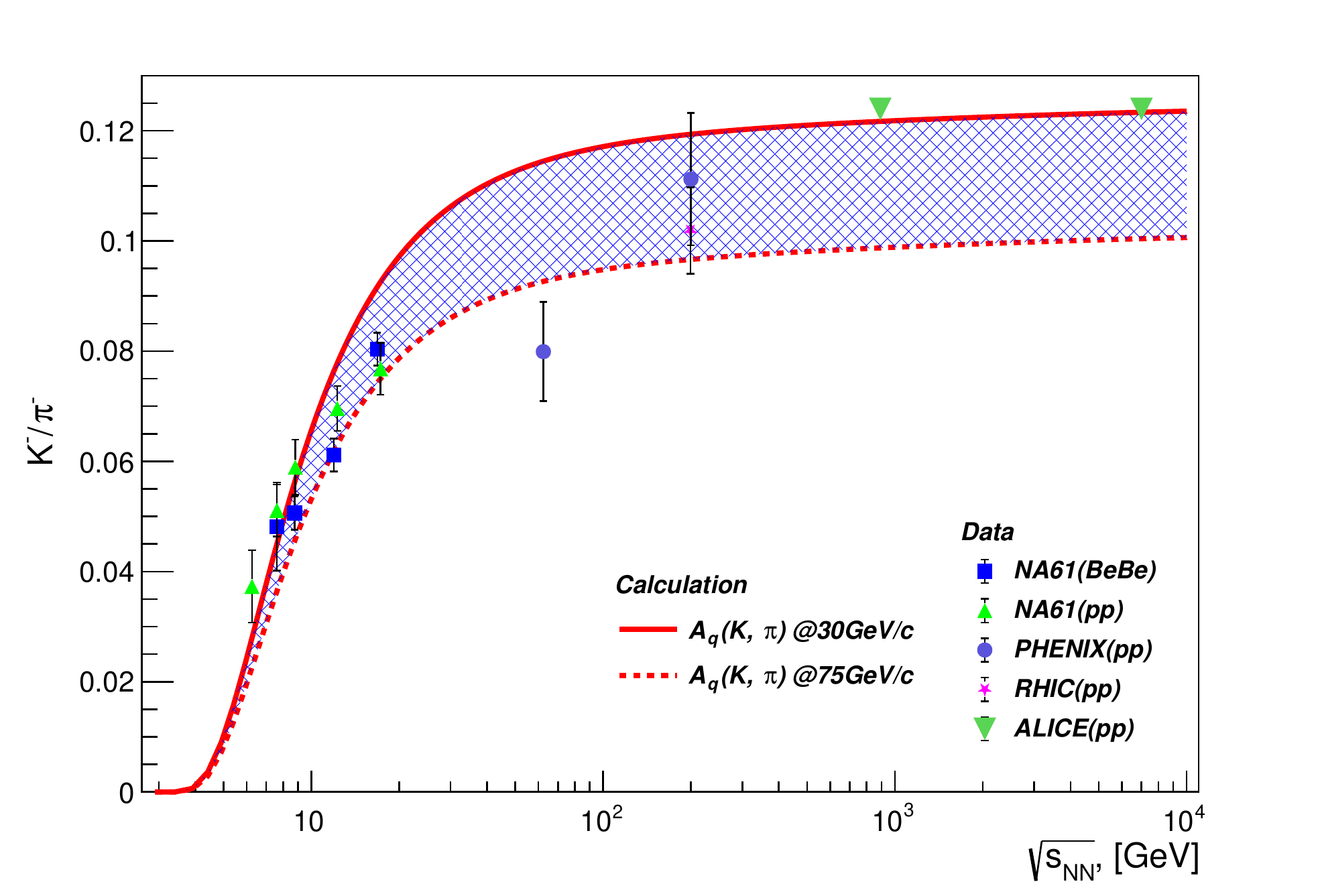}
\end{center}
 \caption{The ratio of $K^-$ and $\pi^-$ meson yields produced in 
$BeBe$(NA61/SHINE) \cite{NA61_BeBe:2021} and $pp$ (NA61/SHINE,PHENIX,STAR,ALICE) 
\cite{NA61_pp:2020,STAR_pp:2011,PHENIX_pp:2011,ALICE_pp:2011,ALICE_pp:2015}  
collisions as a function of $\sqrt{s}$.
}  
\label{fig_ratKmin_20}
\end{figure}
%%%%%%%%%%%%%%%%%%%%%%%%%%%%%%%%%%%%%%%%%%%%%%%%%%%%%%%%%%%%%%%%%%%%%%%%%%%%%%%%%%%%%%%%%%%%%%%%%%%%%%
The upper line in Fig.~\ref{fig_ratKplpipl} corresponds to the fit of data for
$\pi^+$ and $K^+$ mesons at $P_{in}=$ 30$A$ GeV$/$c, and the lower line
corresponds to the similar fit at
$P_{in}=$ 75$A$ GeV$/$c. The upper curve in Fig.~\ref{fig_ratKmin_20} corresponds to the
fit of data for 
$\pi^-$ and $K^-$ mesons at $P_{in}=$ 30$A$ GeV$/$c and the lower line
corresponds to the similar fit at
$P_{in}=$ 75$A$ GeV$/$c.

%{\bf
The $p_T$-spectra of pions and kaons produced in the mid-rapidity of $Be Be$
collisions and their ratios $K/\pi$ as functions of
$\sqrt{s}$ were calculated within different models: Epos 1.99 \cite{Werner:006,Pierog:2018}, 
URQMD 3.4 \cite{URQMD_Bass:1998,URQMD_Bleicher:1999}, AMPT 1.26  \cite{AMPT_Lin:2005,AMPT_Lin:2014,AMPT_Lin:2000},
PHSD 4.00 \cite{PHSD_Cassing:2008,PHSD_Cassing:2009}, SMASH 1.6 \cite{SMASH_Mohs:2020,SMASH_Weil:2016}.
%{\bf
In \cite{NA61_BeBe:2021} 
the comparison of results obtained within these models at the center rapidity region $y=0$
with the NA61/SHINE data was performed. It was illustrated in Figs.(34,35)
of this paper that all these models do not result in the total description of the NA61/SHINE
data at the initial momentum about 150A GeV/c. The energy dependence of the cross sections 
ratio  $K/\pi$ at $y=0$ is described more less satisfactorily within the URQMD model
\cite{URQMD_Bass:1998,URQMD_Bleicher:1999}, see Fig.37 of \cite{NA61_BeBe:2021}.
However, this URQMD model does not describe the $p_T$ and $y$ spectra of pions and kaons 
satisfactorily at the initial momentum about 150A GeV/c , as it is seen from Figs.(34,35) of the NA61/SHINE paper. 
%}
  
\section{Conclusion}

%{\bf
In this paper we have applied the self-similarity approach 
to analyze the 
production of both kaons and pions in $BeBe$ collisions at mid-rapidity 
$y<0.2$ within a wide range of initial energies.
We have presented a self-consistent satisfactory description
of the NA61/SHINE data on $p_T$-spectra of pions and kaons
at $7.62\leq\sqrt{s}\leq 16.84$ GeV.
The fast rise of the $K^+/\pi^+$ and $K^-/\pi^-$ yield ratios as functions of $\sqrt{s}$ 
from the threshold energy of $K^+$ or $K^-$ production up to $\sqrt{s}=$ 20-30 GeV 
has been demonstrated as well as their further slow increase with growing energy.    
The ratio of kaon yields  to
those of pions, produced in $Be Be$ collisions, has been calculated within
this approach as a function of $\sqrt{s}$.

The energy dependence of the ratio of
kaon yields to those of pions in $Be Be$ collision is the same as in $pp$ collision
considered earlier in \cite{LMZ:2021}. The fast
rise with energy of the kaon and pion production  
cross-sections, when $\sqrt{s}$ grows from the threshold energy, is due to the conservation
law of the four-momenta of initial and produced particles and the factor
$\delta_h=1-s^h_{th}/s$ entering into the self-similarity function $\Pi(s,m_{1T},y)$ given by Eq.(7).
The non-zero value of $M$ in the kaon production cross-section results
in the fast rise of the $K^{\pm}/\pi^\pm$ yield ratios 
because in the pion production cross-section $M=$ 0.
When $\sqrt{s}\gg\sqrt{s_{th}}$ and $\sqrt{s}\gg M$,
the pion and kaon production cross-sections and their ratios
become insensitive to factors $\delta_h$ and $M$, however, they are
sensitive to the difference between the quark and gluon contributions to the pion and kaon
spectra as functions of $p_T$ and $\sqrt{s}$. 
That is why the $K^\pm/\pi^\pm$ yield ratios exhibit
two kinds of energy dependence, a fast rise,
when $\sqrt{s_{th}}<\sqrt{s}<$ 20-30 GeV and a slow increase, when 
$\sqrt{s}>$ 20-30 GeV. 

Let us note that no fast rise and no sharp peak in the
ratio between the yields of $K^+$ and 
$\pi^+$ mesons produced in central $BeBe$ collisions are
observed in the NA61/SHINE experiment, according to
\cite{NA61_BeBe:2021}. This ratio is very similar to the same
$K^+/\pi^+$ ratio measured in $pp$ collisions by the
NA61/SHINE Collaboration. 
%}    
%}  
 
\section{Appendix} 
The parameterizations of $\sigma_{tot},\sigma_{SD}$ and $\sigma_{el}$ 
have the following forms \cite{sigma:2013} and \cite{sigm_el:2017} \\
$\sigma_{tot} = (21.7(s/s_0)^{0.0808} + 56.08(s/s_0)^{-0.4525}$) mb;\\ 
$\sigma_{el} = (12.7 - 1.75\mbox{ln}(s/s_0) + 0.14\mbox{ln}^2(s/s_0)$) mb;\\ $\sigma_{SD} = (4.2 +
\mbox{ln}(\sqrt{s/s_0})$) mb.

\begin{figure*}[h!]
	\centering
	\includegraphics[width=6.0in]{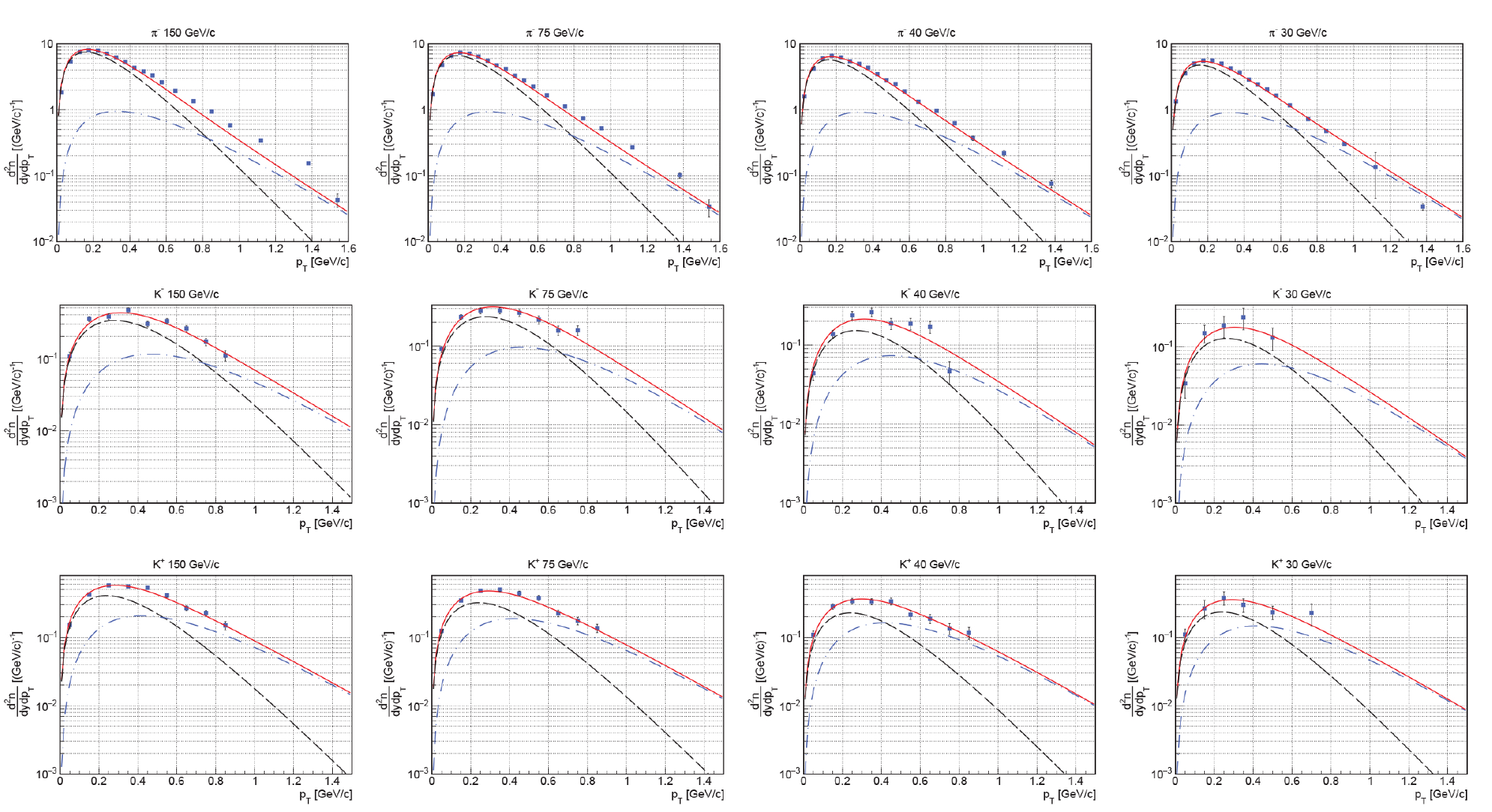}
%{TOTAL_Pt_211020.pdf}
	\caption{The $p_T$ spectra of $\pi^-$, $K^+$ and $K^-$ mesons produced
          at y $\approx$ 0 in inelastic $Be Be$ interactions at SPS energies 
        $\sqrt{s}=$ 7.68 - 16.84 GeV or $P_{in}=$ 30-150 GeV$/$c. 
        The NA61/SHINE data were taken from \cite{NA61_BeBe:2020,NA61_BeBe:2021}.}
	\label{fig_pimin_158}
\end{figure*}

In Fig.~\ref{fig_pimin_158} the $p_T$-spectra of pions and
kaons, produced in the mid-rapidity of $BeBe$ collisions within the
initial momentum range of (30-150) GeV/c, fitted by the
NA61/SHINE data, 
are presented. The black dashed line corresponds
to the quark contribution, the blue dash-dotted curve is the gluon
contribution and the red solid line is the sum of quark and nonperturbative gluon
contributions. The parameters $A_q,A_g$ and $C_q,C_g$ were found from
a fit of NA61/SHINE data and are presented
in Table 1.

As it is shown in \cite{ALM:2015,LM:2018,ML:2020}, the form of
inclusive pion spectra versus $p_T$ at mid-rapidity given by Eqs.~(6-8) 
describes satisfactorily data in a wide range of $\sqrt{s}$ at $p_T<$ 2-3
GeV$/$c. Moreover, as it is shown in \cite{BGLP:2012,GLLZ:2013,LLZ:2014} and
\cite{AJLLM:2018}, the contribution of gluons to the pion
spectrum is related to the gluon
distribution at low $Q^2=$ 1-2 (GeV$/$c)$^2$, the use of which results in 
a satisfactory description of data on hard $pp$ processes at
LHC energies and of proton structure functions at low $x$.
Therefore,  we use Eqs.~(4-7) for the description of data on
pion $p_T$-spectra in $BeBe$ collisions, only  
improving the fit of data. 

As for $K^\pm$ production in $BeBe$ collisions at
not large initial energies
we take into account the additional contribution due to the one Reggeon exchange
diagram, which has a $\sqrt{s_{th}/s}$ dependence. It leads to  
modification of parameter $A_q$ in the following form $A_q(1+\sqrt{s_{th}/s})$, 
which can be approximated by $A_q\mbox{exp}(\sqrt{s_{th}/s})$. This correction 
%$\sqrt{s_{th}/s}$ to $A_q$ 
vanishes at RHIC and LHC energies, however, it allows 
us to describe data at $\sqrt{s}<$ 10 GeV satisfactorily. 

%%%%%%%%%%%%%%%%%%%%%%%%%%%%%%%%%%%%%%%%%%%%%%%%%%%%%%%%%%%%%%%%%%%%%%%%%%%%%%%%%%%%%%%%%%%%%%%%
Parameters $A_q$ for $\pi$, $K^+$ and $K^-$ meson
production were found from the fit of NA61 data
\cite{NA61_BeBe:2021,NA61_BeBe:2020} at initial energies $P_{in}=$ 30$A$-150$A$ GeV$/$c. 
%at $P_{in}=$ 40 GeV, $P_{in}=$ 80 GeV and
%$P_{in}=$ 158 GeV. 
Parameters $A_g$ for $\pi$, $K^+$ and $K^-$ meson
production were found from the fit of NA61 data
at $P_{in}=$ 150$A$ GeV$/$c. 
Other parameters $C_q$ and $C_g$ were taken from fits of NA61 data in
$pp$ collisions.   
%%%%%%%%%%%%%%%%%%%%%%%%%%%%%%%%%%%%%%%%%%%%%%%%%%%%%%%%%%%%%%%%%%%%%%%%%%%%%%%%%%%%%%%
\begin{table*}[h]
\centering
\caption{Table of parameters found from the fit of NA61/SHINE data.}
%Version 3
\begin{tabular}{ | c | c | c | c | c | c | c | c | c | c | c | c| c |}
\hline
Be+Be$ \to h+X$ & \multicolumn{4}{c|}{$\pi^-$} & \multicolumn{4}{c|}{$K^{+}$} & 
\multicolumn{4}{c|}{$K^{-}$}   \\ \hline
$\sqrt{s_{pp}}$, GeV & 16.8 & 11.9 & 8.8 & 7.6 & 16.8 & 11.9 & 8.8 & 7.6 & 16.8 & 11.9 & 8.8 & 7.6 \\ \hline
$\textit{P}$, GeV/$c$ & 150 & 75 & 40 & 30 & 150 & 75 & 40 & 30 & 150 & 75 & 40 & 30 \\ \hline
A$_{q}$ & 17.7 & 15.9 & 13.9 & 11.7 & 3.8 & 3.3 & 2.7 & 3.2 & 9.1 & 7.9 & 7.1 & 7.5 \\ \hline
C$_{q}$ & \multicolumn{4}{c|}{0.147} & \multicolumn{8}{c|}{0.148}\\ \hline
A$_{g}$
& \multicolumn{4}{c|}{6.85} & \multicolumn{8}{c|}{2.963} \\ \hline
C$_{g}$ & \multicolumn{4}{c|}{0.22} & \multicolumn{8}{c|}{0.2271}\\ 
\hline
\end{tabular}
\label{table 1}
\end{table*}
%%%%%%%%%%%%%%%%%%%%%%%%%%%%%%%%%%%%%%%%%%%%%%%%%%%%%%%%%%%%%%%%%%%%%%%%%%%%%%%%%%%%%%%%
%%%%%%%%%%%%%%%%%%%%%%%%%%%%%%%%%%%%%%%%%%%%%%%%%%%%%%%%%%%%%%%%%%%%%%%%%%%%%%%%%%%%%%%%%

{\bf Acknowledgements.}

\begin{sloppypar} 
We are very grateful to K.A. Bugaev, M. Gumberidze, M. Gazdzicki, R. Holzmann,
S. Pulawski, G.Pontecorvo for extremely helpful discussions.   
\end{sloppypar}

%%%%%%%%%%%%%%%%%%%%%%%%%%%%%%%%%%%%%%%%%%%%%%%%%%%%%%%%%%%%%%%%%%%%%%%%%%%%%%%%%%%%%%%%%%%%%

\begin{thebibliography}{00}
\bibitem{NA49:2002}
S.V.~Afanasiev, et al., (NA49 Collaboration) Phys.Rev.C \textbf{66}, 054902 (2002).
\bibitem{NA49:2008}
C.~Alt, et al.,  (NA49 Collaboration) Phys.Rev.C \textbf{77}, 024903 (2008).
%%%%%%%%%%%%%%%%%%%%%%%%%%%%%%%%%%%%%%%%%%%%%%%%%%%%%
\bibitem{Marek:1999}
M.~Gazdzicki, M.I.~Gorenstein, Acta Physika Polon., B {\bf 30}, 2705 (1999).
\bibitem{Marek:2014} 
M.~Gazdzicki, M.I.~Gorenstein, P.~Seyboth, J.Mod.Phys. E {\bf 23}, 1430008 (2014).
\bibitem{Marek:2015}
R.~Poberezhnyuk, M.~Gazdzicki,  M.I.~Gorenstein,  Acta Physika Polon., B {\bf 46}, 1991 (2015). 
\bibitem{Werner:006}
K.~Werner, F.-M. Liu, t. Pirog, Phys.rev. C {\bf 74}, 044902 (2006).
\bibitem{Pierog:2018}
https://web.ikp.kit.edu/rulrich/crmc/html
\bibitem{URQMD_Bass:1998}
S.~Bass, et al., Prog.Part.Nucl.Phys. {\bf 41}, 255 (1998).
\bibitem{URQMD_Bleicher:1999}
M.~Blreicher, et al., J.Phys. G {\bf 25}, 1859 (1999).
\bibitem{AMPT_Lin:2005}
Z.-W.~Lin, et al., Phys.Rev. C {\bf 72}, 06490 (2005).
\bibitem{AMPT_Lin:2014}
Z.-W.~Lin, Phys.Rev. C {\bf 90}, 014904 (2014).
\bibitem{AMPT_Lin:2000}
B.~Zhang,  et al., Phys.Rev. C {\bf 61}, 067901 (2000).
\bibitem{PHSD_Cassing:2008}
W.~Cassing, E.L.~Bratkovskaya, Phys.Rev. C {\bf 78}, 034919 (2008).
\bibitem{PHSD_Cassing:2009}
W.~Cassing, E.L.~Bratkovskaya, Nucl.Phys. A {\bf 831}, 215 (2009).
\bibitem{SMASH_Mohs:2020}
J.~Mohs:, S.~Ryu, H.~Elfner, J.Phys. G {\bf 47}, 065101 (2020).
\bibitem{SMASH_Weil:2016}
J.~Weil, et al., Phys.Rev. C {\bf 94}, 054905 (2016).
%%%%%%%%%%%%%%%%%%%%%%%%%%%%%%%%%%%%%%%%%%%%%%%%%%%%%
\bibitem{NA61_BeBe:2021}
A.~Acharya, et al., (NA61/SHINE Collaboration) Eur. Phy. j. C {\bf 81}, 73
(2021).
\bibitem{NA61_ArSc_ratio:2021}
 Maja Mackowiak-Pawłowska for the NA61/SHINE Collaboration, arXiv:2112.01877 [nucl-ex]. 
\bibitem{NA61_ArSc:2021}
A.~Acharya \textit{et al.}, (NA61/SHINE Collaboration), Eur. Phys. J. C \textbf{81}, 397 (2021)
\bibitem{NA61/SHINE:2020} 
A.~Aduszkiewicz, et al., (NA49 Collaboration) Phys.Rev.C \textbf{102}, 011901(R) (2020).
\bibitem{LMZ:2021}
G.I.~Lykasov, A.I.~Malakhov, A.A.~Zaitsev, Eur. Phys. J. A {\bf 57 }, 78 (2021).
\bibitem{c3}  
{A.M.~Baldin, L.A.~Didenko}, Fortsch.Phys. \textbf{38}, 261 (1990).
\bibitem{c4}  
{A.M.~Baldin, A.I.~Malakhov, and A. N.~Sissakian}, Phys. Part. Nucl. \textbf{29} (Suppl. 1), 4 (2001).
\bibitem{4}
A. M.Baldin, A. A. Baldin. Phys. Particles and Nuclei, {\bf 29} No3, 232 (1998).
\bibitem{5}
A.M.~Baldin, A.I.~Malakhov. JINR Rapid Communications, No.1(87)-98, pp.5-12
(1998).
\bibitem{Baldin_AA:1996}
 Baldin  A.A.  JINR  Rapid  Comm. No. 4[78]-96  p.61-68.
\bibitem{Fermi:1950}
{E.~Fermi}, Phys. Rev. \textbf{92}, 452 (1953)
\bibitem{Pomeran:1951}
{I. Ya.~Pomeranchuk}, Izv. Dokl. Akad. Nauk Ser.Fiz. \textbf{78}, 889 (1951).
\bibitem{Landau:1953}
{L.D.~Landau}, Izv. Akad. Nauk Ser. Fiz. \textbf{17}, 51 (1953).
\bibitem{Hagedorn:1965} 
{R.~Hagedorn}, Supplemento al Nuovo Cimento \textbf{3}, 147 (1965).
%\bibitem{Hagedorn:1973}
%{R.~Hagedorn, L.~Montvay}, Nucl.Phys. B \textbf{59}, 45 (1973).
%\bibitem{Hagedorn:1983}
% {R.~Hagedorn}, Rivista del Nuovo Cimento \textbf{6}, \textnumero{10}, 1 (1983).
%%%%%%%%%%%%%%%%%%%%%%%%%%%%%%%%%%%%%%%%%
\bibitem{ALM:2015}
D.A.~Artemenkov, G.I.~Lykasov, A.I.~Malakhov,
Int.J.Mod.Phys. \textbf{A30}, 1550127 (2015)
\bibitem{LM:2018}
G.I.~Lykasov, A.I.~Malakhov, Eur. Phys. J. A {\bf 54}, 187 (2018).
\bibitem{ML:2020}
A.I.~Malakhov, G.I.~Lykasov,  Eur. Phys. J. A {\bf 56}, 114 (2020).
\bibitem{AJLLM:2018} {A.M.~Abdulov, H.~Jung, A.V.~Lipatov, G.I.~Lykasov, M.A.~Malyshev},
 Phys.Rev. \textbf{D98}, 054010 (2018).
%%%%%%%%%%%%%%%%%%%%%%%%%%%%%%%%%%%%%%%%%%%%%%%%%%%%%%%%%%%%%%%%%%%%%%%%%%
\bibitem{BGLP:2012}
V.A.~Bednyakov, A.A.~Grinyuk, G.I.~Lykasov, M.~Pogosyan,
Int.J.Mod.Phys., \textbf{A27}, 1250042 (2012).  
\bibitem{GLLZ:2013} {A.A.~Grinyuk, G.I.~Lykasov, A.V.~Lipatov, N.P.~Zotov}, 
Phys.Rev. \textbf{D87}, 074017 (2013).
\bibitem{LLZ:2014} {A.V.~Lipatov, G.I.~Lykasov, N.P.~Zotov},
 Phys.Rev. \textbf{D89}, 014001 (2014).
\bibitem{AJLLM:2018} {A.M.~Abdulov, H.~Jung, A.V.~Lipatov, G.I.~Lykasov, M.A.~Malyshev},
 Phys.Rev. \textbf{D98}, 054010 (2018).
\bibitem{sigma:2013}
N.~Cartiglia, arXiv:1305.6131 [hep-ex].
\bibitem{sigm_el:2017}
S.H.~Stark, Eur.Phys.J. (Web of Conf.) {\bf 141} 03007 (2017).   
\bibitem{Thermmod:1993}
E.~Schnedermann, J.~Sollfrank, U.~Heinz, Phys.Rev.C48,2462 (1993).
\bibitem{Wilk:2000}
G.~Wilk, Z.~Wlodarczyk, Phys.Lett. {\bf 84}, 2770 (2000).
\bibitem{Bugaev1:2002}
K.A.~Bugaev, J.Phys.G:Nucl.Phys., {\bf 28}, 1981 (2002).
\bibitem{Bugaev2:2002}
K.A.~Bugaev, M.~Gadzicki, M.I.~Gorenstein, Phys.Lett. B{\bf 544}, 127 (2002). 
\bibitem{Cleymans:2013}
J.~Cleymans, G.I.~Lykasov, A.S.~Parvan, et al., Phys.Lett. B {\bf 723}, 351 (2013).
\bibitem{7}
J.L.~Kley, et al., E895 Collaboration, Phys.Rev. C {\bf 68}, 054905 (2003).
%\bibitem{11}
%J.~Cleymans, J.~Struempfer, L.~Tirko, Phys.Rev. C {\bf 78}, 017901 (2008).
\bibitem{10} 
N.~Abgrall, et al., NA61/SHINE Collaboration, Eur. Phys. J. C {\bf 74}, 2794 (2014).
\bibitem{c11} {K.A.~Ter-Martirosyan}, Sov.J.Nucl.Phys., \textbf{44}, (1986) 817.
%\bibitem{c13} {A.A.~Grinyuk, G.I.~Lykasov, A.V.~Lipatov, N.P.~Zotov}, 
%Phys.Rev. \textbf{D87}, (2013) 074017. 
\bibitem{NA61_pp:2020}
A.Adiszkiewicz, \textit{et al.}, (NA61/SHINE Collaboration) 
Phys.Rev.C \textbf{102} 1, 011901 (2020).
\bibitem{STAR_pp:2011}
B. I. Abelev \textit{et al}. (STAR Collaboration)
Phys. Rev. C \textbf{79}, 034909 (2009).
\bibitem{PHENIX_pp:2011}
A. Adare \textit{et al}. (PHENIX Collaboration)
Phys. Rev. C \textbf{83}, 064903 (2011).
\bibitem{ALICE_pp:2011}
Aamodt K. \textit{et al}. (ALICE Collaboration)
Eur. Phys. J. C \textbf{71}, 1655 (2011).
\bibitem{ALICE_pp:2015}
J. Adam \textit{et al}. (ALICE Collaboration)
Eur. Phys. J. C \textbf{75}, 226 (2015).
\bibitem{NA61_BeBe:2020}
A. Acharya \textit{et al}.,(NA61/SHINE Collaboration), Eur. Phys. J. C \textbf{80}, 961 (2020).
\end{thebibliography}
\end{document}